# Multilayer InSe-Te van der Waals heterostructures with ultrahigh rectification ratio and ultrasensitive photoresponse

Fanglu Qin[†], Feng Gao[‡], Mingjin Dai[‡], Yunxia Hu[‡], Miaomiao Yu[†], Lifeng Wang[§], PingAn Hu[*,‡], Wei Feng[*,†]

[†] School of Chemistry, Chemical Engineering and Resource Utilization, Northeast Forestry University, Harbin, 150040, China

[‡] Key Lab of Microsystem and Microstructure of Ministry of Education, Harbin Institute of Technology, Harbin, 150080, China

[§] Institute for Frontier Materials, Deakin University, 75 Pigdons Road, Waurn Ponds, Geelong, Victoria 3216, Australia

*Email: wfeng@nefu.edu.cn, hupa@hit.edu.cn

**Abstract:** Multilayer van der Waals (vdWs) semiconductors have great promising application in high-performance optoelectronic devices. However, the photoconductive photodetectors based on layered semiconductors often suffer from large dark current and high external driven bias voltage. Here, we report a vertical van der Waals heterostructures (vdWHs) consisting of multilayer indium selenide (InSe) and tellurium (Te). The multilayer InSe-Te vdWHs device shows a record high forward rectification ratio greater than $10^7$ at room temperature. Furthermore, an ultrasensitive and broadband photoresponse photodetector is achieved by the vdWHs device with an ultrahigh photo/dark current ratio over $10^4$, a high detectivity of $10^{13}$, and a comparable responsivity of 0.45 A/W under visible light illumination with weak incident power. Moreover, the vdWHs device has a photovoltaic effect and can function as a self-powered photodetector (SPPD). The SPPD is also ultrasensitive to the broadband spectra ranging from 300 nm to 1000 nm and is capable of detecting weak light signals. This work offers an opportunity to develop next-generation electronic and optoelectronic devices based on multilayer vdWs structures.

## Introduction

Photodetectors, converting the optical signals into the electrical signals, have enormous applications in communication, national defense, and health monitoring. To develop high-performance photodetectors, various nanomaterials have been investigated, such as zero-dimensional (0D) quantum dots[1-3] and one dimensional (1D) nanowires[4-8]. Recently, two-dimensional (2D) van der Waals (vdWs) semiconductors have gained increasing attention and have been explored as the candidates of the high-performance photodetectors due to their excellent optical and electrical properties[9-11]. Many photodetectors based on vdWs semiconductors have been demonstrated. Monolayer $MoS_2$ photodetector shows a high responsivity of 880 A/W to visible light[12]. Multilayer GaSe and GaS photodetectors have a good response to ultraviolet (UV) light (254 nm) on rigid[13] and flexible substrates[14]. Few-layer black phosphorus (BP) is sensitive to near-infrared (NIR) with a high photoresponsivity[15, 16]. Among the van der Waals semiconductors, InSe is particularly captivating. Multilayer InSe has a small direct bandgap of 1.26 eV, a high absorption coefficient, and excellent electronic properties[17]. The multilayer InSe photodetectors show the high-performance, broadband photoresponse ranging from UV to IR[18-20]. However, similar to other photoconductive photodetectors, the multilayer InSe photodetectors inevitably suffer from a high driven bias voltage (10 V)[18] and large dark current (μA)[19].

The designing a p-n junction is a good way to overcome those drawbacks. The p-n junction is an important device since it is a foundation block for photodetector, diode, and solar cell. Several p-n junctions based on multilayer InSe have been developed[21-24]. The $CuInSe_2$-InSe lateral p-n heterostructure shows a good photoresponse to UV-vis[21]. Multilayer InSe-GaTe van der Waals heterojunctions (vdWHs) show a high-performance and fast response to 405 nm light with zero external bias, demonstrating that InSe based p-n junction is promising for application in photodetector with a high speed, low power consumption, and low dark current[23]. The photodetector based on InSe-BP diode shows a broadband (450-900 nm) response with a highly polarization-sensitive photocurrent[24]. However, BP is extremely sensitive and unstable in air, hampering its practical application[25].

Tellurium (Te), an elemental semiconductor with p-type transport behavior, has drawn global attention due to its excellent electrical properties and good air-stability[26]. Physical vapor deposition[27] and hydrothermal method[28] have been successfully applied to synthesis multilayer Te nanosheets. High-performance field-effect transistors (FETs) based on multilayer Te have been demonstrated with a high hole mobility of 700 $cm^2V^{-1}s^{-1}$ and large current on/off ratio of $10^6$, which are comparable to that of BP[29] and are superior to that of other 2D p-type materials, such as $WSe_2$ device[30] and GaTe device[31]. Moreover, good air-stability makes 2D Te more attractive than BP for practical applications. The multilayer Te photodetector shows high photoresponse to short-wave infrared due to its small bandgap[32]. However, the multilayer Te based heterojunction is lacking.

Here, we report a vdWHs device that can work as a high-performance forward diode, photodetector, and self-powered photodetector (SPPD). The vdWHs consist of multilayer InSe and Te nanosheets. The InSe-Te vdWHs diode achieves an ultrahigh forward rectification ratio exceeding $10^7$, which is the record value for forward diodes made of multilayer vdWHs devices. More importantly, the vdWHs device can work as an ultrasensitive and broadband photoresponse photodetector with an ultrahigh photo/dark current ratio over $10^4$, a high detectivity of $10^{13}$, and a comparable responsivity of 0.45 A/W under visible light illumination with weak incident power. Moreover, the vdWHs device shows a photovoltaic effect and can work as an SPPD. The SPPD is also ultrasensitive to the broadband spectra ranging from 300 nm to 1000 nm and is capable of detecting weak light signals (nW).

## Results and Discussion

The fabrication processes of vertical multilayer InSe-Te vdWHs device were presented in supporting information. Figure 1a is the 3D schematic of vertical multilayer InSe-Te vdWHs device, and multilayer InSe locates on top of multilayer Te. Figure 1b is a typical optical image of InSe-Te vdWHs device. The thickness of multilayer InSe and Te nanosheets are 42 nm and 120 nm, respectively, which are determined by atomic force microscopy (AFM) as shown in Figure S1. The quality of isolated InSe, Te, and heterojunction is evaluated by the Raman spectrum as shown in Figure 1c. For individual InSe, three Raman peaks are locating at 114, 177, and 226 cm$^{-1}$, ascribing to $A^1_{1g}$, $E^1_{2g}$, and $A^2_{1g}$ modes, respectively, which agree well with early reports for InSe nanosheets[19]. There Raman peaks are locating at 92, 120 and 140 cm$^{-1}$, corresponding to $E_1$-TO, $A^1$, and $E^1$ modes, respectively, agreeing well with previous reports of multilayer Te nanosheets[28, 32]. Both multilayer InSe and Te show three Raman peaks, which agree well with early reports. The heterojunction clearly shows both InSe and Te characteristic Raman peaks as same as the Raman spectra of isolated InSe and Te, suggesting high quality in the junction area. Notably, the intensities of Raman peaks of InSe nanosheets are increased at heterojunction, this interesting behavior should be further investigated in the future. To further investigate the structure of InSe-Te vdWHs, scanning electron microscope (SEM) and energy dispersive spectra (EDS) mapping are measured as displayed in Figure 1d. Uniform element distributions are observed in the heterojunction area, further demonstrating successfully build InSe-Te vdWHs.

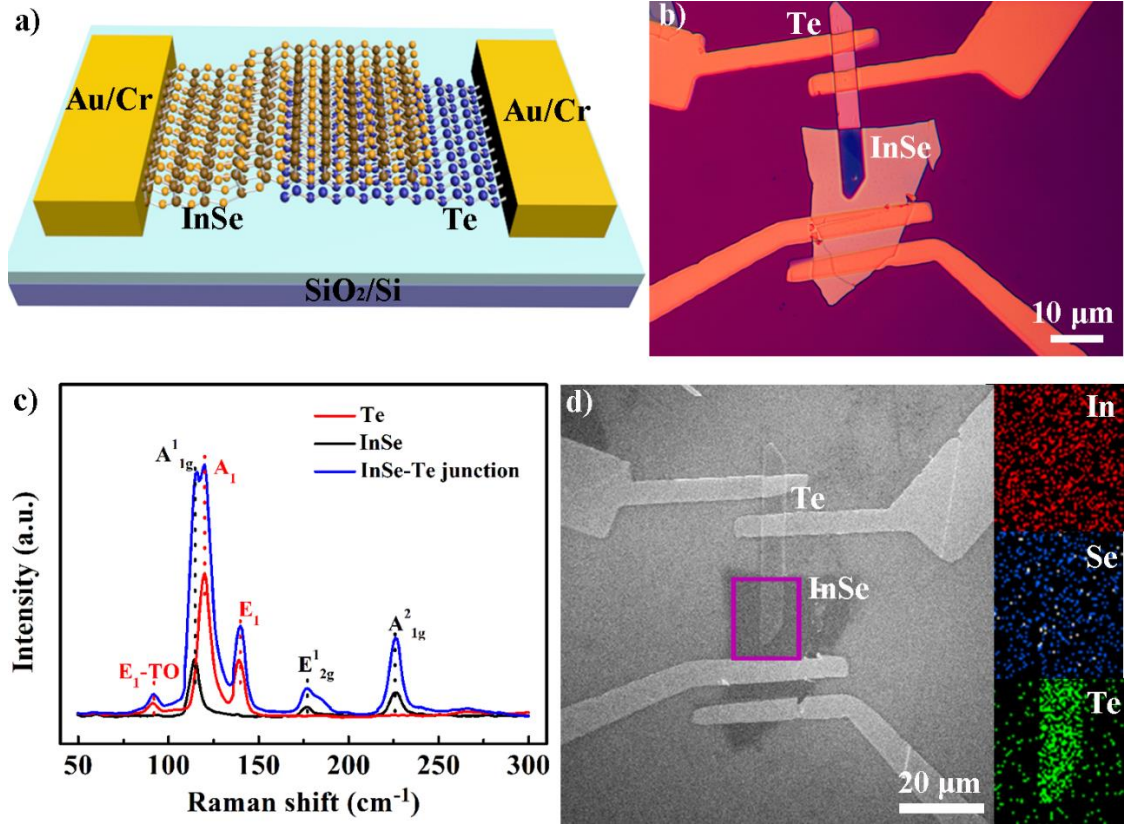

**Figure 1.** Structure characterizations of InSe-Te vdWHs: a) Schematic of InSe-Te vdWHs. b) An optical image of InSe-Te heterojunction device. c) Raman spectra of multilayer InSe, Te and the heterojunction areas. d) The SEM of InSe-Te vdWHs and corresponding EDS maps of devices.

We first characterized the electrical transport properties of individual multilayer InSe and Te nanosheets before we investigated the electrical properties of InSe-Te heterojunction. Figure 2a shows the transfer curves of isolated multilayer InSe and Te nanosheets measured at $V_{ds}$ = 1 V, where the blue line represents InSe nanosheet and the black line is Te nanosheet, respectively. Multilayer InSe and Te nanosheets show typical n-type and p-type electronic transport behaviors, respectively, which agree well with early reports of multilayer InSe and Te nanosheets[18, 28]. Figure S2 shows the $I$-$V$ curves of isolated InSe and Te nanosheet without gate voltage ($V_g$). The current linearly depends on bias voltage, indicating the ohmic contact is formed between InSe-Cr and Te-Cr, respectively. Figure 2b shows an ideal energy band structure of InSe-Te vdWHs, which is a typical type-II band alignment. Therefore, the multilayer InSe-Te vdWHs device is predicted to function as a forward p-n diode. Then the electrical properties of multilayer InSe-Te vdWHs were measured. Figure 2c is the $I$-$V$ curves of multilayer InSe-Te vdWHs measured at a small bias voltage ($V_{ds}$) range of $V_{ds} = \pm 3$ V with $V_g = 10$ V. The multilayer InSe-Te vdWHs device shows a strong current rectifying behavior with a large forward current (3.7 μA) and an ultralow reverse current (lower than pA), demonstrating that a forward diode is built successfully. The forward rectification ratio value is calculated by the ratio of the forward current to reverse current. The multilayer InSe-Te vdWHs diode shows an ultrahigh rectification ratio greater than $10^7$ at $V_g = 10$ V, which is the record value for forward diodes made of multilayer vdWHs devices as depicted in Figure 2d[21-24, 33-39]. The $I$-$V$ curve of the vdWHs device is totally different from those of isolated multilayer InSe and Te (Figure S2), indicating that the strong current-rectifying features of InSe-Te vdWHs diode are mainly owed to well-structured vdWs p-n junction rather than schottky barrier. The $I$–$V$ curves of the InSe-Te vdWHs diode are measured under various gate voltage ($V_g$) and a gate-dependent rectifying behavior is observed as shown in Figure S3. The forward current increases and reverse current decreases as the $V_g$ increases from -10 V to 10 V, leading to the rectification ratio increases from $10^4$ to $10^7$. The ultrahigh forward rectification ratio suggests that the multilayer InSe-Te vdWHs device can function as a high-performance forward diode. The transfer curve of multilayer InSe-Te vdWHs is measured. As shown in Figure S4, the vdWHs device shows an n-type transport behavior, suggesting that the transport properties of the InSe-Te vdWHs device mainly depends on the multilayer InSe channel. The current on/off ratio is $10^2$, which is much lower than $10^7$ of isolated multilayer InSe. This is attributed to a strong screening effect from underling thick Te layer and the gate-modulation on upper layer InSe becomes weaker.

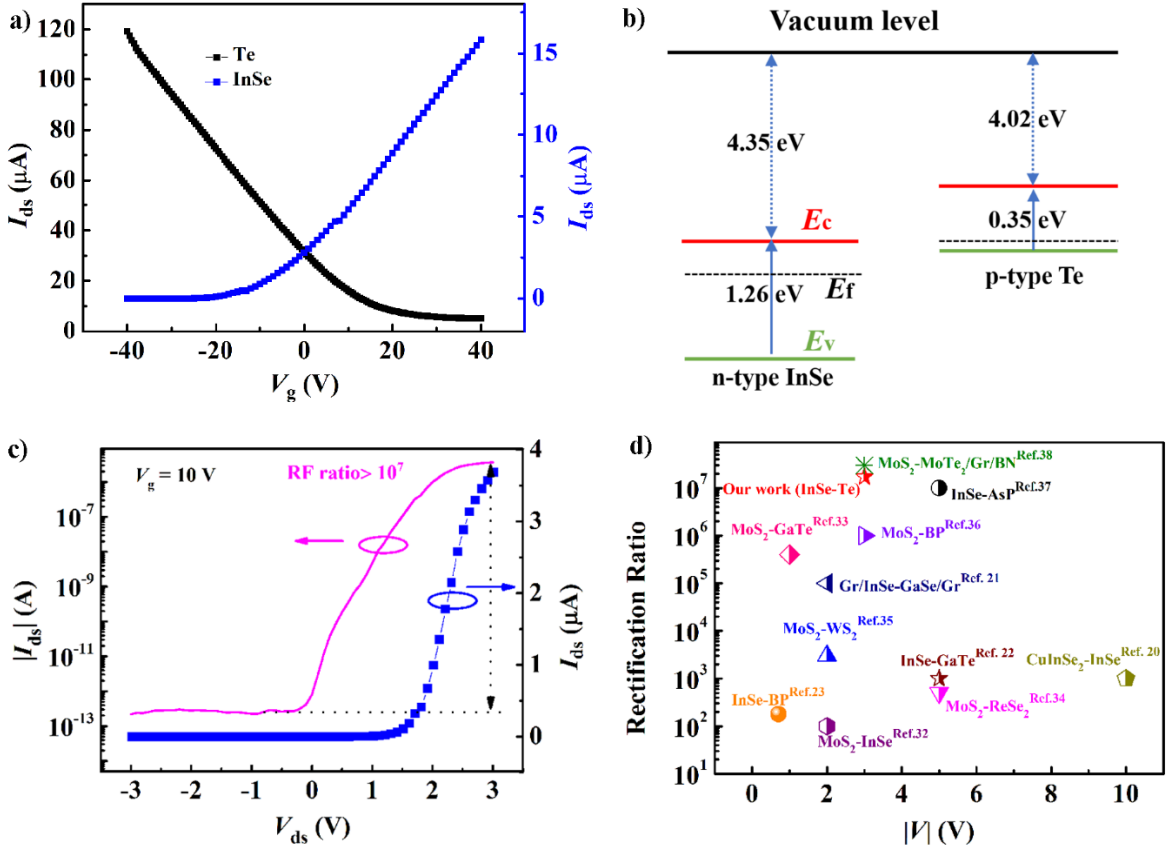

**Figure 2**. Electrical properties of InSe-Te vdWHs device. a) Transfer cures of individual multilayer InSe and Te nanosheets. b) Energy band structure of InSe-Te heterojunction. c) The $I-V$ curves of InSe-Te vdWHs device at $V_g = 10$ V. d) Comparison of rectification ratio of various diodes based on multilayer vdWHs devices measured at room temperature. Notes: All the diodes are forward diodes except ref. 37 and 38 listed in the Figure 2d.

Next, we characterize the photoresponse properties of InSe-Te vdWHs as photodetector and photovoltaic devices. In this study, all the photoresponse properties of vdWHs device were measured at $V_g = 0$ V unless otherwise specified. Various wavelength lights vertically irradiate on our device. Figure 3a is the $I-V$ curves of the InSe-Te vdWHs device irradiated by various wavelength lights with weak incident powers (nW). The multilayer InSe-Te vdWHs device shows a broadband photoresponse ranging from UV (300 nm) to NIR (1000 nm). The multilayer InSe is highly resistive (Figure 2 and S4) at $V_g = 0$ V, suggesting that most of the external bias voltage is applied to multilayer InSe. Therefore, the photocurrent ($I_{ph}$, $I_{ph} = I_{illumination}-I_d$) mainly stems from the photogenerated carriers in upper multilayer InSe nanosheets drove to the external circuit.

Hence, the vdWHs devices show a broadband photoresponse due to benefited from a small bandgap of multilayer InSe (1.26 eV). The photo/dark current ratio can be calculated by $I_{ph}/I_d$, where $I_d$ is dark current. Figure 3b and S5 show the photo/dark current ratio of various wavelength lights at $V$ = -2 V. Notable, the ultrahigh photo/dark current ratios of $10^3$ - $10^4$ are observed in a broadband spectra ranging from UV to NIR with weak incident powers (~nW), which are attributed to the ultralow dark current (~0.1 pA) and the good optical absorption in multilayer InSe[18]. The ultrahigh photo/dark current ratio is one or two magnitudes higher than most of the vdWHs devices (more details in Table S1). The vdWHs device shows the better photoresponse to 400 nm light because of photo/dark current ratio shows a peak at 400 nm, which is consistent with light absorption spectrum of multilayer InSe[40]. Hence, in this study we mainly focus on photoresponse performance of the multilayer InSe-Te vdWHs device illuminated by 400 nm. To further investigate the signal-to-noise performance of the vdWHs device, the linear dynamic range (LDR) is calculated by the equation: LDR = $20\log(I_{ph}/I_d)$. The calculated LDR is 93.6 dB for 400 nm light at $V$ = -2 V, which is much higher than 66 dB of commercial InGaAs photodetector.

For further investigate the photoresponse properties, the vdWHs device is irradiated under 400 nm light with various incident powers and the $I$-$V$ curves are depicted in Figure 3b. As shown in Figure 3a inset, the InSe-Te vdWHs device shows an obvious positive open-circuit voltage ($V_{oc}$) and negative short-circuit current ($I_{sc}$) under various wavelength lights illumination, suggesting that it can function as a photovoltaic device. The electrical power ($P_{el}$) can be calculated by $P_{el}$ = $V_{ds} \times I_{ds}$ and $P_{el}$ *vs* incident power under 400 nm light is plotted in Figure S6a. The $P_{el}$, $I_{sc}$, and $V_{oc}$ increase as the incident power increases as shown in Figure S6. The power conversion efficiency (PCE), defined as $\eta_{pv}$ = $P_{el,\,m}/P_{in}$ (where $P_{in}$ is incident power), is calculated to be 1.26% under 400 nm light with a weak incident power of 14 nW, which is comparable and even higher than most of the multilayer vdWHs based photovoltaic devices.[23]

To further appraise the photoresponse properties of the vdWHs device as a photodetector, three important figure of merits are extracted (more calculated details in Supporting Information): responsivity ($R$), detectivity ($D*$) and external quantum efficiency (EQE). Figure 3d and S7 shows the calculated $R$, $D*$, and EQE values of the InSe-Te vdWHs device under 400 nm light with various incident powers at $V$ = -2 V. The calculated $R$, $D*$, and EQE values are 0.45 A/W, $1.1 \times 10^{13}$ Jones, and 136.5% under the incident power of 1.7 nW, respectively, which are not only comparable to or higher than commercial Si and InGaAs photodetectors (1 A/W, $10^{12}$ Jones), are also superior to most of the vdWHs devices based photodetectors (find more details in Table S1). The calculated $R$ values decrease with increasing incident powers, which is a common phenomenon in photodetectors based on multilayer van der Waals materials. This incident power-dependent $R$ mainly originates from defects and traps at the interface between InSe/Te and substrates[18]. For practical application, photoswitching stability and photoresponse speed are another two key figure of merits for a high-performance photodetector. Figure 3e is the multicycles $I$-$T$ curves measured under various wavelength lights on and off. The current of our device rises rapidly and maintains a stable state as various wavelength lights are on, and decays quickly when various lights are off, demonstrating that the multilayer InSe-Te vdWHs device has good stability and repeatable photoswitching behavior for the broadband spectra. To obtain the photoresponse speed accurately, fast photoswitching behavior is measured and recorded. Figure 3f is the fast photoswitching illuminated by 532 nm laser at $V$ = -3 V. The response speed is defined as the time for the $I_{ph}$ increasing from 10% to 90% for rise process and the $I_{ph}$ decreasing from 90% to 10% for decay process. The response speeds are 600 μs and 800 μs for the rise and decay processes, respectively, indicating that the InSe-Te vdWHs device has a fast response speed. As mentioned above, the ultrahigh photo/dark current ratio, ultrahigh LDR, high detectivity, comparable responsivity, and fast response speed indicate that the multilayer InSe-Te vdWHs device has great potential application in high-performance photodetector.

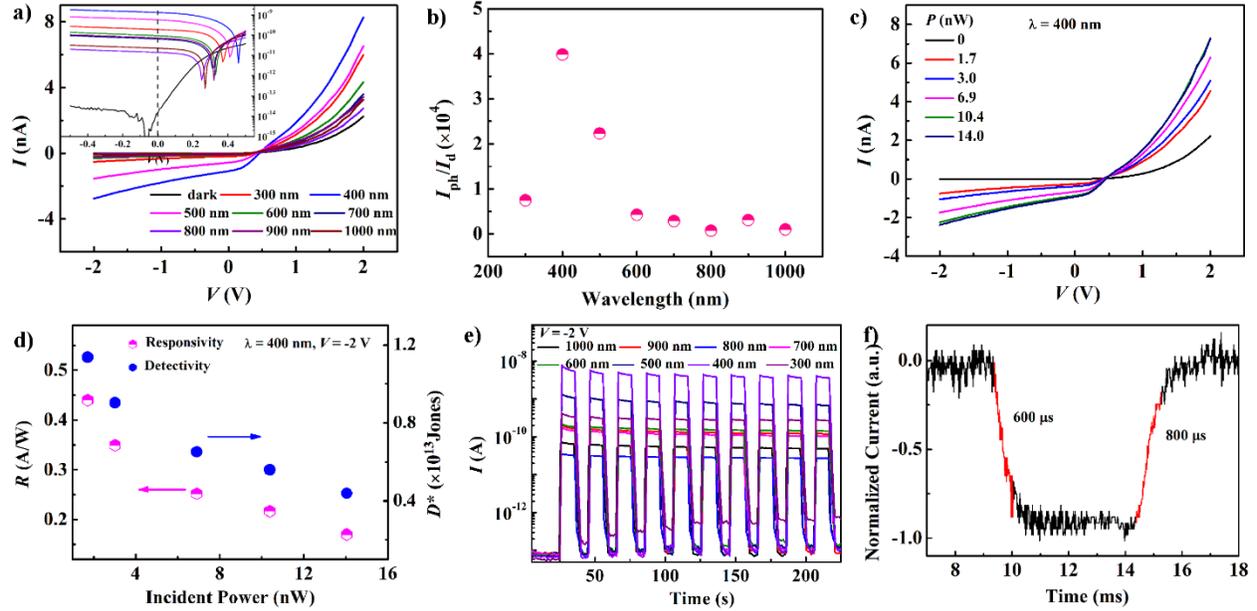

**Figure 3**. Photoresponse properties of the InSe-Te vdWHs device. (a) The *I-V* curves of the vdWHs device illuminated by various lights. The inset is *I-V* curves measured at small bias range (-0.5~0.5 V). The incident powers are 3.34, 10.37, 13.33, 11.44, 8.37, 2.85, 27.75, and 27.32 nW for 300, 400, 500, 600, 700, 800, 900, and 1000 nm, respectively. (b) The photo/dark current ratio under various wavelength lights at *V* = -2 V. (c) The *I-V* curves of the InSe-Te vdWHs device are irradiated by 400 nm light with various incident powers. (d) The calculated *R* and *D\** values irradiated by 400 nm light at *V* = -2 V. (d) The Time-dependent photoresponse of the InSe-Te vdWHs device irradiated by various lights at *V* = -2 V. (f) Response speed of the InSe-Te vdWHs device irradiated by 532 nm laser at *V* = -2 V.

Benefiting from the photovoltaic effect, we can explore the photoresponse properties of the multilayer InSe-Te vdWHs device as an SPPD. The SPPD also has a broadband photoresponse ranging from UV (300 nm) to NIR (1000 nm) as shown in Figure 3a inset. The photo/dark current ratio is greater than $10^4$ and $10^3$ under visible lights and NIR illumination with the weak incident powers in Figure 4a, respectively, suggesting that the SPPD is ultrasensitive to weak light signals in a broadband spectra. Figure 4b is calculated *R* and *D\** for 400 nm light with various incident powers. The $R$ is 170 mA/W for 400 nm light, which is greater than most of the vdWHs devices based SPPD (find more details in Table S1). The $D^*$ values are exceeding $10^{12}$ Jones for 400 nm light, respectively, which are comparable to that of commercial Si and InGaAs based photodetectors. The SPPD shows a stable photoswitching behavior under various wavelength lights (300-1000 nm) as shown in Figure 4c. The photoresponse times of SPPD are 900 and 950 μs for rise process and decay process, respectively, which are a little larger than that of the vdWHs device operated at $V$ = -2 V. The slower response speed of SPPD is due to the much smaller built-in electric field, leading to a longer transit time for the photogenerated carriers in SPPD to arrive the external circuit. All the above results suggest that our InSe-Te vdWHs device can function as an SPPD.

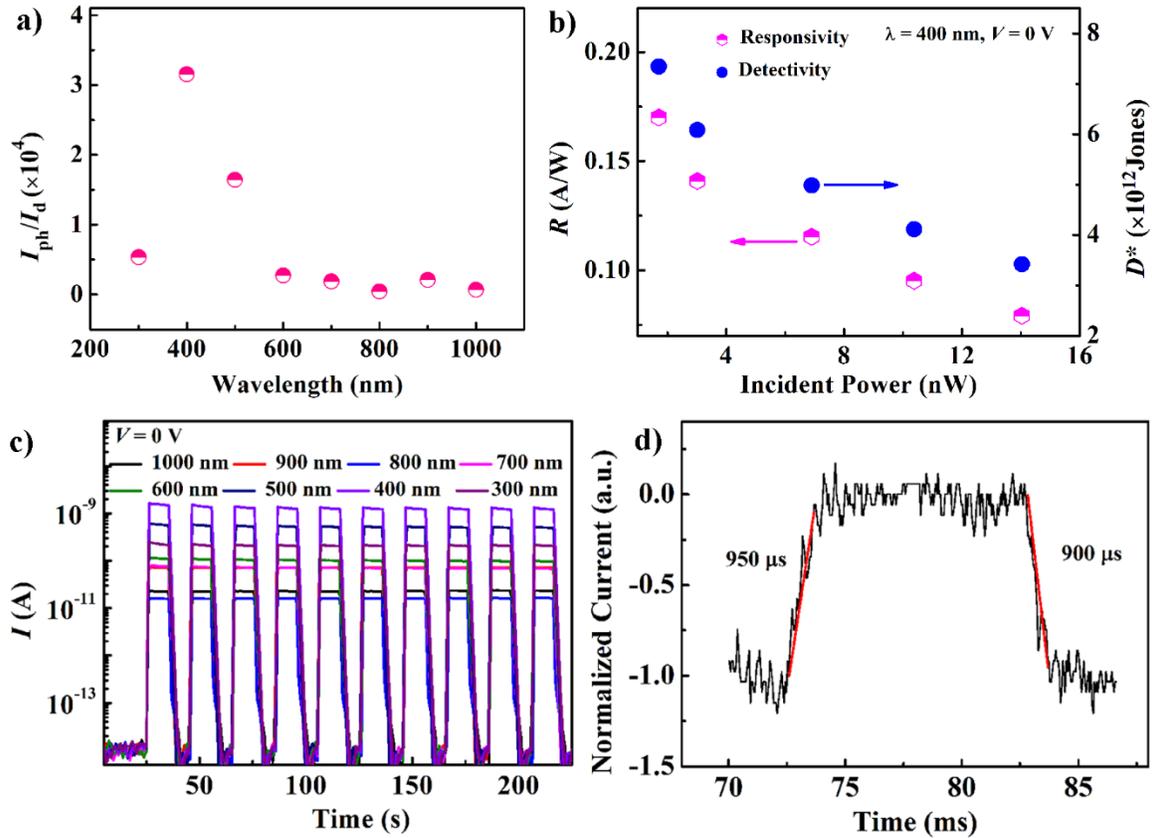

**Figure 4**. Self-powered photodetector based on InSe-Te vdWHs. (a) The photo/dark current ratio under various wavelength lights at $V$ = 0 V. (b) The calculated $R$ and $D^*$ values irradiated by 400 nm light at $V$ = 0 V. (c) The Time-dependent photoresponse of the InSe-Te vdWHs device irradiated by various lights at $V = 0$ V.

## Conclusions

In summary, we have successfully fabricated the multilayer InSe-Te vdWHs device and demonstrate that the device can work as a high-performance forward diode and photodetector. Notably, an ultrahigh forward rectification ratio greater than $10^7$ is achieved. Moreover, the high-performance photodetector is demonstrated with an ultrahigh photo/dark current ratio over $10^4$, a high detectivity of $10^{13}$, and a comparable responsivity of 0.45 A/W under visible light illumination with weak incident power. Furthermore, the InSe-Te vdWHs device also has a photovoltaic effect and can work as an SPPD. The SPPD is also ultrasensitive to the broadband spectra ranging from 300 nm to 1000 nm and is capable of detecting weak light signals. Our work offers an opportunity to develop next-generation electronic and optoelectronic devices based on multilayer vdWs structures.

## ASSOCIATED CONTENT

**Supporting Information**. The Supporting Information is available free of charge on the ACS Publications website. Descriptions of the experiment, calculated equations for responsivity, detectivity and external quantum efficiency.

## AUTHOR INFORMATION

**Corresponding Author**

*Email: wfeng@nefu.edu.cn.

*Email: hupa@hit.edu.cn.

**Notes**

The authors declare no competing financial interest.

## Acknowledgments

We gratefully acknowledge the support from the National Natural Science Foundation of China (NSFC, No. 51802038), the China Postdoctoral Science Foundation funded project (No. 2019T120246 and 2018M630329) and Heilongjiang Postdoctoral Special Fund (No. LBH-TZ1801).